\documentclass[9pt,twocolumn,twoside]{optica}
\usepackage{amsmath}
\usepackage{hyperref}
\usepackage[sort&compress]{natbib}
\usepackage{url}
\setboolean{shortarticle}{false}
\setboolean{minireview}{false}

\title{Temporal shaping and time varying orbital angular momentum of displaced vortices.}
\author[1]{Fazele Hosseini}
\author[1,*]{Mohammad A. Sadeghzadeh}
\author[1,2]{Amir Rahmani}
\author[3,4]{Fabrice P. Laussy}
\author[5]{Lorenzo Dominici}

\affil[1]{Department of Physics, Yazd University, Yazd, Iran}
\affil[2]{Department of Physics, Azarbaijan Shahid Madani University, Tabriz, Iran}
\affil[3]{Faculty of Science and Engineering, University of Wolverhampton, Wulfruna St, Wolverhampton WV1 1LY, UK}
\affil[4]{Russian Quantum Center, Novaya 100, 143025 Skolkovo, Moscow Region, Russia}
\affil[5]{	CNR NANOTEC, Istituto di Nanotecnologia, Via Monteroni, 73100 Lecce, Italy}

\affil[*]{msadeghzadeh@yazd.ac.ir }

\begin{abstract}
  The fundamental mode of rotation in quantum fluids is given by a
  vortex, whose quantized value yields the orbital angular momentum
  (OAM) per particle. If the vortex is displaced (off-centered) from
  the reference point for rotation, the angular momentum is reduced
  and becomes fractional.  Such displaced vortices can further exhibit
  a peculiar dynamics in presence of confining potentials or couplings
  to other fields.  We study analytically a number of 2D systems where
  displaced vortices exhibit a noteworthy dynamics, including
  time-varying self-sustained oscillation of the OAM, complex
  reshaping of their morphology with possible creation of
  vortex-antivortex pairs and peculiar trajectories for the vortex
  core with sequences of strong accelerations and decelerations which
  can even send the core to infinity and bring it back. Interestingly,
  these do not have to occur conjointly, with complex time dynamics of
  the vortex core and/or their wavepacket morphology possibly taking
  place without affecting the total OAM. Our results generalize to
  simple and fundamental systems a phenomenology recently reported
  with Rabi-coupled bosonic fields, showing their wider relevance and
  opening prospects for new types of control and structuring of the
  angular momentum of light and/or quantum fluids.
\end{abstract}

\setboolean{displaycopyright}{true}

\begin{document}

\maketitle

\section{Introduction}

Topology is essential to different areas of physics, in particular to
condensed matter physics~\cite{stan17,cast17}.  This involves
invariants under deformation from the ground state to a new
configuration for which any path to the ground state is energetically
too costly. As such, topological invariants powered by Homotopy
groups provide subtle tools to
detect holes in a given space. Examples are topological defects in
systems with continuous broken symmetry. A vortex is a typical such
defect in systems with global~$U(1)$ broken symmetry. It is
characterized by a null density region and an integer (topological)
charge associated to a singularity in the gradient of the phase of the
order parameter, which defines the core of the vortex. In this way,
the topological charge (TC) defines an intrinsic orbital angular
momentum (OAM) embedded in the vortex state.  Vortices are common in
many media and systems, including light~\cite{dennis09a,shen19a} and
quantum matter, such as superconductors~\cite{blatter_vortices_1994},
superfluids~\cite{leggett_superfluidity_1999} or atomic Bose Einstein
condensates~\cite{Matthews1999}, among others. Vortices also recently
flourished in coupled light-matter, so-called ``polaritons'',
systems~\cite{Lagoudakis2008,lagoudakis09a,sanvitto10b,nardin11a,roumpos11a,dominici15b,dominici18a}
that can interpolate between the purely optical and strongly
interacting condensed matter cases.

A vortex beam (VB)---a vortex confined in a beam with a spiral
wavefront---is special as it can carry angular
momentum~\cite{yao2011}, which can further be distinguished between
one of two kinds: either with a non-uniform phase-varying wavefront,
which is called an ``anisotropic VB'', or with a uniform phase
variation, which is referred to as an ``isotropic VB''. In both cases,
an interesting situation arises when the core of the vortex is
displaced from the center of the beam, a situation which does not
arise in infinite-size systems. This leads to a reduced angular
momentum, which is still fixed, and to a modified morphology for the
phase of the field (its distribution in space~\cite{yao2011}). For a
static VB (not varying in time), either isotropic or anisotropic, it
is known that the vortex angular momentum can be described by a
combination of elementary vortices with integer topological charges,
which determines the overall morphology and associated angular
momentum across the beam~\cite{ROUX04}.

Vortex beams can also be dynamic, i.e., their angular momentum can be
time dependent. It is only recently, however, that VB have entered
this regime~\cite{domicini18a,Rego2019a}. The earliest report, to the
best of our knowledge, was by some of the present authors and
collaborators~\cite{domicini18a} in the highly versatile polaritonic
platform~\cite{kavokin17a}. There, a time-varying angular momentum was
shown to be self-sustained by the interplay between Rabi oscillations
and VB morphologies in two coupled fields. This was achieved by
preparing the system in a special topological initial condition
imprinted in the microcavity polariton field with two delayed pulses
of different topological charge. This produced a rich temporally and
spatially structured dynamics of the off-centered core in the
polariton fluid resulting in oscillating linear and angular momenta of
the emitted light. Such oscillations happen in the linear regime and
can be described exactly. We study further this case
in the last Section of the present text. A similar scheme, with two
retarded pulses but now partially overlapping in time, has also been
used in the nonlinear process of high harmonic generation, resulting
in a vortex beam with continuously time-varying angular
momentum~\cite{Rego2019a}.

Here, we study analytically fundamental and simple physical systems
that display similar nontrivial motion of a dynamical vortex. The
results are quite remarkable given the complex and far-reaching
phenomenology that is displayed by a simple theoretical model, which
basically reduces to a first-year quantum mechanics textbook
exposition. Given the ease of access of both the experimental scheme
(combinations of pulsed excitations with various TC) and the variety
of possible platforms where to explore such Physics, we expect the
field of dynamic VB to quickly take off in a variety of systems in the
near future. What we find is that such displaced vortices sustain a
rich dynamics with a morphology that is stretched or folded up in time
to accompany the variation of the angular momentum. Interestingly,
although one may expect a time-varying morphology to also result in a
time-varying angular momentum, we show that this is not compulsorily
the case. To this end, we consider a displaced vortex in several
elementary systems, namely, an infinite (circular) quantum well
potential (Section~\ref{se:fdjgiugheiur}), a slightly anharmonic
potential (Section~\ref{sec:kdfg78t53nejnr}), a spatially squeezed
harmonic potential (Section~\ref{sec:dbcfweghweywesq}) and the already
discussed case of two coupled condensates (polaritons, in
Section~\ref{sec:jerh98392nf948}).  Section~\ref{sec:erk8347yrfn347y}
concludes and gives general remarks. It is worth noting that all the
examples considered hereon are in the linear regime, where the
self-interaction between particles is negligible. The phenomenology is
however robust in the sense that it is maintained in the presence of
nonlinearities or interactions, as we have checked numerically. In
such cases, however, further complications arise with more complex
reshaping of the vortices, proliferations of vortex-antivortex pairs
and other features which we leave for other studies.

% We consider in  the case of displaced vortex
% in quantum well with hard boundary, for which we introduce the
% mechanism of pair excitement induced exclusively by quantum effects
% associated with interference. While the vortex morphology is
% peculiarly varied in time, the angular momentum content remains
% steady. 

%  accounts for the off center
% vortex in the subtle example of harmonic potential, where we extend
% the study to consider the effect of the spatial anisotropic
% anharmonicity (deviation from harmonic potential) and time--harmonic
% perturbation (oscillatory in time) with a gain/dissipative term. We
% show that while the harmonic potential alone sustains an angular
% momentum which is constant, on the other hand the time harmonic
% perturbation induces periodically varying angular
% momentum. Interestingly, anharmonicity modifies greatly the morphology
% of the vortex alongside with the vortex core motion, as removing the
% degeneracy of the excited states of harmonic potential mediates
% oscillatory time--varying angular momentum. 

% The other possibility to observe the time--varying angular momentum is
% through binary condensates, namely, Rabi coupled fields, which is
% studied in . Such behavior was shown recently for the case of large
% imbalance between masses of coupled fields, corresponding to the
% microcavity polaritons.  
%
\section{Infinite circular quantum well}\label{se:fdjgiugheiur}

Our first platform to consider the dynamics of a
  displaced-vortex in a VB is the fully-confined case with the
  canonical circular geometry, namely, the infinite circular quantum
  well.  This provides an example of interactions between vortices and
  the system's boundaries. As we show below, interference between the
  rotating wavepacket and its reflection from the hard walls leads to
  a time-dependent morphology of the phase, which, nevertheless,
  leaves the expectation value for the total angular momentum
  invariant, as expected from the conservation
  rules. Manipulating the size of the initial packet
    introduces different content of radial momentum adding up to the
    rotational one. This leads to various regimes for the dynamics,
  ranging from a smooth rotation of the pattern to pair-creation of
  vortices/antivortices. Such a case can be realized
    experimentally for instance with polaritons in etched
    micropillars~\cite{Abdal18,Luko18,Sedov19}. The Hamiltonian for
a circular quantum well of radius $R$ reads as
\begin{align}
H_{qw}=-\frac{\hbar^2}{2m}\nabla^2+V(r)\,,
\end{align}
with
\begin{align}
V(r)=
\begin{cases}
0&r<R\\
\infty&r>R
\end{cases}\,,
\end{align}
where $r=\sqrt{x^2+y^2}$ and $\varphi=\arg(x+iy)$. The system is
easily integrated exactly. Solving the equation
$H_{qw}\chi_{n,l}=E_{n,l}\chi_{n,l}$, one finds the eigenstates and
eigenenergies as
\begin{subequations}
	\begin{align}
	\chi_{n,l}&=\frac{N_{n,l}}{\sqrt{2\pi}}e^{il\varphi}J_{|l|}(k_{n,l}r)\,,\\
	E_{n,l}&=\frac{\hbar^2 k_{n,l}^2}{2m}=\frac{\hbar^2}{2mR^2}\beta_{n,l}^2\,,
	\end{align}
\end{subequations}
where $J_{|l|}$ is the Bessel function of integer order, $N_{n,l}$ is
the normalization constant, $l=0,\pm1,\pm2,\cdots$, and
$\beta_{n,l}=k_{n,l}R$ is the $n$th zero of the $|l|$th Bessel
function for which $J_{|l|}(\beta_{n,l})=0$ as imposed by the boundary
condition, namely $\chi_{n.l}(R,\varphi)=0$. From these stationary
solutions $\chi_{n,l}$, one then solves Schr\"odinger's equation
$i\hbar\partial_t\psi=H\psi$ for any initial
state~$\psi_0(r,\varphi)\equiv\psi(r,\varphi,0)$ as the linear
superposition
\begin{subequations}
	\begin{align}\label{eq:urut8758hrerbwe}
	\psi(r,\varphi,t)&=\sum_{n,l}\alpha_{n,l}e^{-iE_{n,l}t/\hbar}\chi_{n,l}(r,\varphi)\,,\\
	\alpha_{n,l}&=\int d\varphi \int rdr\psi_0\chi_{n,l}^\ast(r,\varphi)\,.
	\end{align}\
\end{subequations}
To describe the dynamics of a displaced vortex, we
  take for the initial condition a superposition of two Bessel--Gauss
  wavepackets, one with topological charge $\text{TC}=1$ and another
  with $\text{TC}=0$:
	\begin{align}
	\label{eq:jsdhfjd896w978rfr}
	\psi_0=\begin{cases}
	e^{-r^2/2w^2}(J_1(k_{1,1}r)e^{i\varphi}+a_cJ_0(k_{1,0}r))&r<R\\
	0&r>R
	\end{cases}\,,
	\end{align}
	where $w$ is the spot size of the wavepacket and $a_c$ is a
        free parameter that upon tuning the relative strength between the two wavepackets controls the offset position of the vortex core. The
      wavepacket solution in time is then given by
\begin{align}\label{eq:fhgioerhjg0utrjopier}
  \psi(r,\varphi,t)=\sum_n \bigg[\alpha_{n,1}e^{-iE_{n,1}t/\hbar}\chi_{n,1}+ \alpha_{n,0}e^{-iE_{n,0}t/\hbar}\chi_{n,0}\bigg]\,.
\end{align}
\begin{figure*}[t]
  \begin{center}
    \includegraphics[width=\linewidth]{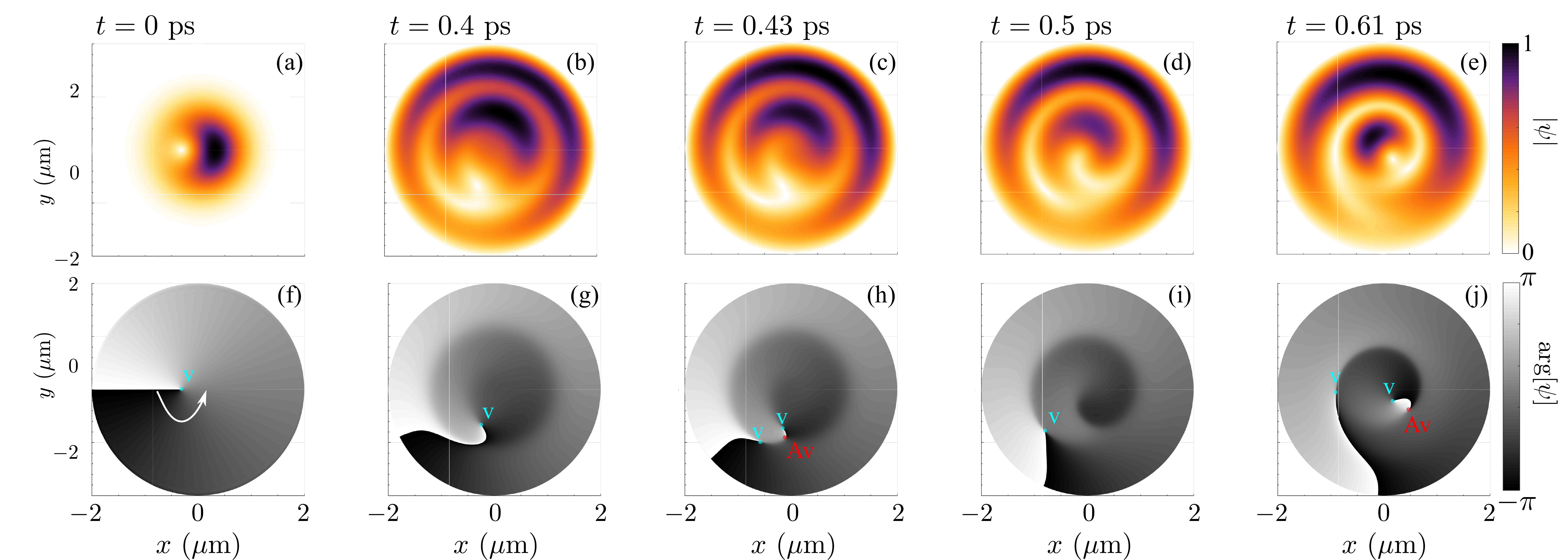}
    \caption{Dynamics of an off-centered (displaced) vortex in an
      infinite circular quantum well, in the regime~$w\ll R$. The
      upper panels show the density profile~$|\psi|$ of the wavepacket
      at different times. The lower panels illustrate the
      corresponding phase map $\arg[\psi]$. In (h), the phase gets so
      distorted that a pair is excited, shown by V (vortex) and AV
      (antivortex). The pair moves until V and AV recombine. It gets
      excited again later, as shown in (j). Parameters:
      $R=2~\mu\mathrm{ m}$, $w=0.5~\mu\mathrm{ m}$, and $a_c=0.3$.  }
    \label{fig:rytuwteitrgywey89382hj}
  \end{center}
\end{figure*}  

The dynamics strongly depends on the size~$w$ of the
  packet. When $w$ is of the order of quantum well size~$R$, the terms
  in Eq.~(\ref{eq:fhgioerhjg0utrjopier}) that dominate are those
  weighted by $\alpha_{1,1}$ and $\alpha_{1,0}$, which results in a
  smooth rotation of the wave packet, with simply a quantum beating
  between two states. However, when $w$ is smaller than $R$ by one
  order of magnitude or more, a larger number $\alpha_{n,i}$
($i=0,1$) enter the dynamics, which acquires a markedly different
character as higher~$n$ amplitudes become significant.  Examples are
shown in Fig.~\ref{fig:rytuwteitrgywey89382hj}. We can observe
different stages in the dynamics. At first, we see an increase of the
wavepacket size inside the well as a result of its diffusion, which
also separates the minima and maxima of the density. However, since
the wavepacket should remain zero at the boundary of the well, there
is a point in time where diffusion stops and a second crest starts to
form, with a half-moon shape, being thinner than the first one but
larger in radius. At the same time, the vortex core itself keeps
rotating (Fig.~\ref{fig:rytuwteitrgywey89382hj}g) and the distortion
of the phase excites new vortices, namely, a vortex-antivortex pair to
conserve angular momentum
(Fig.~\ref{fig:rytuwteitrgywey89382hj}h). These are moving inside the
fluid until their later recombination
(Fig.~\ref{fig:rytuwteitrgywey89382hj}i). Shortly
  after that, another vortex-antivortex pair is formed locally
  (Fig.~\ref{fig:rytuwteitrgywey89382hj}j). Such processes are
repeated during the full time evolution of the dynamics (See also the Supplementary Movie SM1~\cite{Movie1} for better
  illustration of the dynamics). Despite the complex motion of the
core inside this fluid bouncing back and forth as it rotates inside
the well, the angular momentum remains steady. This is expected since
the external potential is a purely confining one, being radially
symmetric and azimuthally homogeneous. Considering the potential
gradient and its symmetry, its boundaries act both as a local and net
force on the fluid at any moment, however there is no net torque
acting on the fluid.  Therefore, while the center of mass of the fluid
and its net linear momentum keep changing due to the continuous
bouncing of the fluid against the well boundaries, its net angular
momentum remains constant. Yet, some peculiar morphology reshaping,
involving also the topological charges, happens during the evolution.
It is interesting to check how this is self-consistently ensured by
the formalism which produces correspondingly intricate dynamics of the
fields:
\begin{align}
  \label{eq:Thu30Jul125731CEST2020}
\langle L_z\rangle=-i\hbar\int d\varphi\int r dr \psi^\ast\partial_\varphi\psi=\frac{\hbar R^2}{2}\sum_n |\alpha_{n,1}|^2N_{n,1}^2\big[J_2(\beta_{n,1})\big]^2\,.
\end{align}
Although the way the Bessel functions balance each others in space is
not transparent, one can see indeed how
Eq.~(\ref{eq:Thu30Jul125731CEST2020}) produces a time-independent
average. 

Another interesting quantity is the orbital angular momentum per particle,
which in the case above is less than one, as expected for an
off-center vortex~\cite{Maji19}, namely,
$\langle \tilde{L}_z\rangle\equiv \langle L_z\rangle/N<1$, with $N$ the total number of
particles~\footnote{Also, in the next sections we refer to $\langle \tilde{L}_z\rangle$ as the mean (expectational) value of orbital angular momentum per particle.}, given by
\begin{align}
N=\frac{R^2}{2}\bigg[ \sum_n |\alpha_{n,1}|^2N_{n,1}^2\big[J_2(\beta_{n,1})\big]^2+ \sum_m |\alpha_{m,0}|^2N_{m,0}^2\big[J_1(\beta_{m,0})\big]^2\bigg]
\end{align}
The condition $\langle \tilde{L}_z\rangle<1$ comes from the initial
field~(\ref{eq:fhgioerhjg0utrjopier}), where we have introduced two kinds of
particles with a fraction only of them carrying (integer and equal to
one) angular momentum. The ratio of particles that carry angular
momentum is responsible for the position of the core, making it closer
to the center (less displaced) as more particles have angular
momentum (i.e., for smaller $a_c$). Similarly, increasing the number of particles without
angular momentum (i.e., for larger $a_c$) pushes the core towards the boundary.

\begin{figure}[tb]
	\begin{center}
          \includegraphics[width=\linewidth]{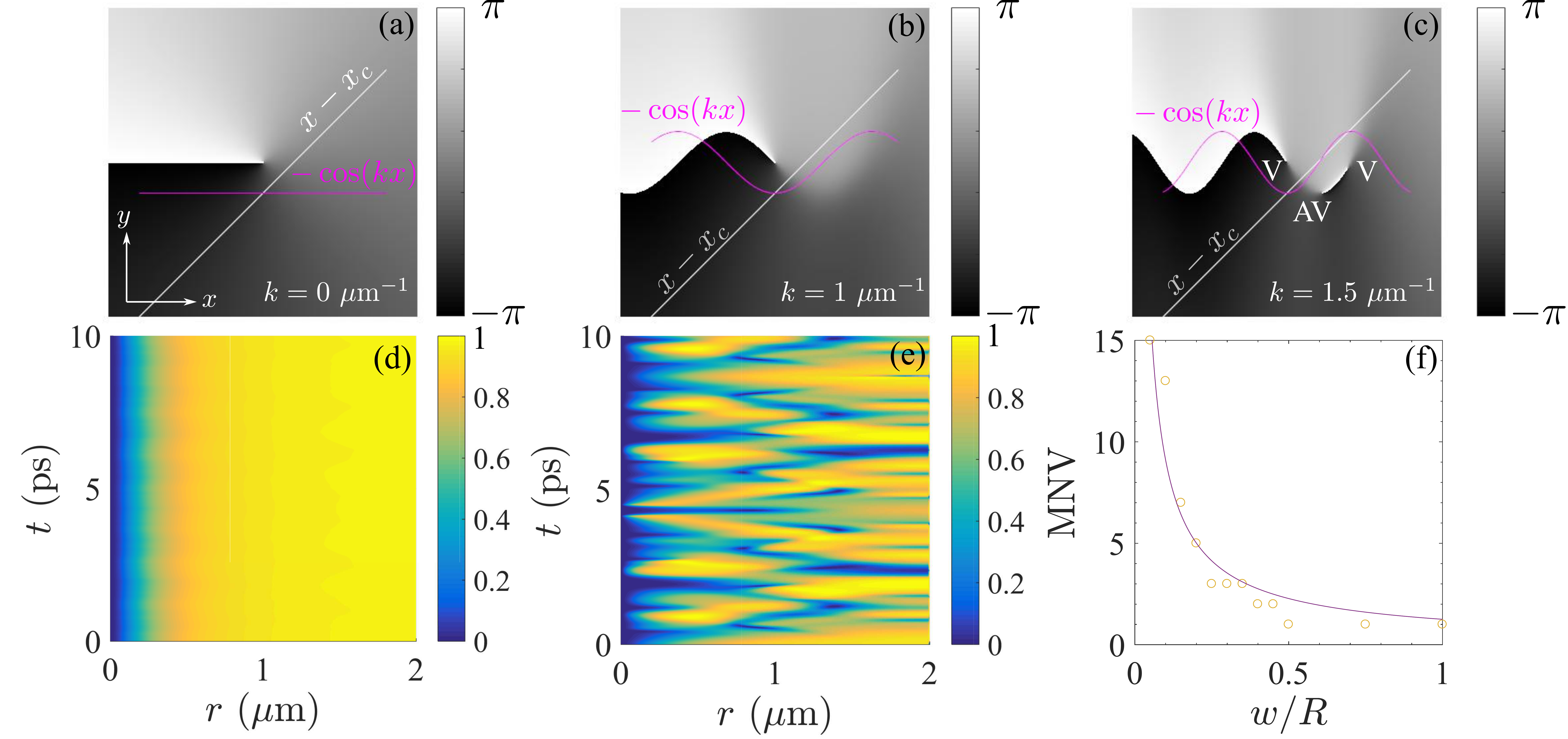}
          \caption{Upper panels: toy-model for vortex-antivortex pair
            creation, caused by the phase distortion from the
            interference of two waves, one being oblique with respect
            to the other and carrying a topological charge
            $\text{TC}\leq1$. The frames show different $k$, with
            increasing distortion of the phase. In (a) and (b) there
            is only one root, corresponding to the intersection of two
            curves $x-x_c$ (shown in white) and $-\cos(kx)$ (shown in
            magenta). In (c) there are three roots, bringing an
            additional vortex (V) and antivortex~(AV). Panels (d) and
            (e) show the radial OAM per particle (defined in
            Eq.~(\ref{eq:d37t273gd2edg7832})) that quantifies angular
            momentum locally, showing how the orbital angular
            momentum, always constant in time, is also locally
            constant for large wavepackets, as in~(d) with
            $w=5~\mu\mathrm{m}$, $a_c=0.3$ and $R=2~\mu\mathrm{m}$,
            but displays a rich local time-dependence of the OAM, as
            in~(e) with $w=0.5~\mu\mathrm{m}$. Panel (f) shows the
            maximum number of vortices (MNV) observed during the
            dynamics for different values of the wavepacket size
            $w$. The solid line is a (negative power law) fitting curve highlight
            the nonlinear dependence of MNV on the initial size
            $w$. Here we used $a_c=0.2$ and $R=2~\mu\mathrm{m}$.}
          \label{fig:rytuwasateitrgywey89382hj}
	\end{center}
\end{figure}

Although both the total and mean value of the angular
  momentum is constant for the whole wavepacket, one can consider a
  local orbital angular momentum, which reveals time-varying
  features. In particular, we can define
  \begin{equation}
    \langle L_z\rangle_r\equiv-i\hbar\int_0^r\int_0^{2\pi}
    \rho\psi^\ast(\rho,\varphi,t)[\partial_\varphi\psi](\rho,\varphi,t)d\varphi\,d\rho
  \end{equation}
  the mean angular momentum enclosed \emph{in} a circle of radius~$r$,
  with
  $\langle L_z\rangle=\lim_{r\rightarrow\infty}\langle
  L_z\rangle_r$. Then the mean angular momentum \emph{on} a cirle of
  radius~$r$ is obtained as $d\langle L_z\rangle_r/dr$. Similarly, one
  can define the mean number of particles on a circle of radius~$r$ as
  $dN_r/dr$ with
  $N_r=\int_0^r\int_0^{2\pi}
  \rho|\psi(\rho,\varphi,t)|^2d\varphi\,d\rho$, so that the chain rule
  yields the mean value of OAM per particle on a circle of
  radius~$\rho$, that we call~$\langle \tilde{L}_z\rangle_r$, as
  $\langle \tilde{L}_z\rangle_r=\frac{d\langle L_z\rangle_\rho}{d\rho}{d\rho\over dN_\rho}$, which
  is thus given by:
  \begin{align}\label{eq:d37t273gd2edg7832}
    \langle \tilde{L}_z\rangle_r=-i\hbar\frac{\int_{0}^{2\pi}d\varphi \psi^\ast(r,\varphi,t)[\partial_\varphi\psi](r,\varphi,t)}{\int_0^{2\pi}d\varphi |\psi(r,\varphi,t)|^2}\,.
  \end{align}
  This quantity is shown in panels d-e of
  Fig.~\ref{fig:rytuwasateitrgywey89382hj} where it is seen how the
  local OAM exhibits strikingly different behaviours for different
  cases. For large $w$, with a smooth rotation of the field, the local
  angular momentum is constant in time on any radial distance from the
  center of the quantum well. However, by decreasing the initial
  packet size, which induces a nontrivial dynamics of the vortex, one
  then finds strongly time-varying features of the angular momentum
  but only locally, since these cancel when averaging over the entire
  wavepackets.  This is shown in panel~(e). At a given radius $r$, a
  torque can be felt locally due to the internal dynamics of the
  field, which is absent in panel (d). One can see that variations are
  stronger in the central part rather than close to the boundaries
  ($r=0, 2$), where the density of the field has less dynamics. This
  is clear from panels (c-d) of Fig.~\ref{fig:rytuwteitrgywey89382hj},
  where a half-moon shaped density near the boundary remains almost
  steady while the central part has a rich dynamics of vortex rotation
  and pair excitation.

The vortex-antivortex pair formation, evolution and later
recombination, that all occur without affecting the total angular
momentum, but that can be detected locally, can be understood from the
expanding cloud reaching the boundary and being reflected, producing a
circular ripple of lower density and with a larger phase gradient
(visible starting from Fig.~\ref{fig:rytuwteitrgywey89382hj}b,g
on). This interference between the outward diffusing and the inward
reflected waves is the origin to the secondary vortex-antivortex pair,
which is nucleated starting from this loop of locally low density. The
pair creation can be modelled by using a toy-model which interferes
two waves, both with a Laguerre Gaussian type envelope of same
amplitude and propagating in the directions $\mathbf{k}_1$ and
$\mathbf{k}_2\neq\mathbf{k}_1$ as
$A_1=e^{-i\mathbf{k}_1\cdot\mathbf{r}}e^{-r^2/(2w^2)}$ and
$\text{TC}=0$,
$A_2=e^{-i\mathbf{k}_2\cdot\mathbf{r}}e^{-r^2/(2w^2)}\big(re^{i\varphi}-r_ce^{i\varphi_c}\big)$,
with a TC. If, for simplicity, $A_1$ travels along the $x$ direction
($\mathbf{k}_1=k\hat{\imath}$) and $A_2$ does not propagate
($\mathbf{k}_2=0$), then the two interfering waves yield a total
density $I\propto |e^{ik x}+re^{i\varphi}-r_ce^{i\varphi_c}|$.  The
condition for a vanishing density is given as $-\cos(kx)=x-x_c$ and
$-\sin(kx)=y-y_c$. Depending on the wavenumber $k$, such a density can
have either one or more roots.  The former case
corresponds to the initial topological charge $\text{TC}\leq1$, while when there are three roots, a vortex-antivortex
  pair is added to the initial TC. An example for this behavior of
the phase is shown in Fig.~\ref{fig:rytuwasateitrgywey89382hj} a-c,
where we take $x_c=1~\mu\mathrm{ m}$ and $y_c = 0$ (obtained for the
polar parameters $r_c = 1~\mu\mathrm{m}$ and $\varphi_c = 0$). One can
see how, upon increasing $k$, the phase gets distorted to the point of
creating a vortex pair.  This condition also depends on the relative
amplitude of the two waves, which we did not consider to highlight the
importance of the momentum. Coming back to the displaced vortex in the
circular quantum well, the excitation of the pair comes into play when
the reflected wave from the hard boundary interferes with the field
inside the well, matching the toy-model above, where a
vortex-antivortex pair is created and annihilated repeatedly. One
could also consider higher-energy initial conditions with a larger
number of vortex-antivortex pairs. The maximum number
  of vortices (MNV) observed during the dynamics as a function of the
  initial packet size $w$ is shown in
  Fig.~\ref{fig:rytuwasateitrgywey89382hj}f.  It is clear that by
  decreasing $w$, MNV is growing in a nonlinear
  fashion. Interestingly, also configurations of even number of
  vortices are possible, where a single vortex or antivortex can be
  excited from the boundary of the quantum well. In fact, since the
  distance of a vortex fractionalises its contribution to the OAM, and
  a pair of vortex/antivortex immediately drift apart with different
  distances, it is clear that the field itself must locally
  accommodate for the corresponding changing angular momentum, and it
  is thus not surprising that, would a vortex come from the boundary
  of the well, it is allowed to drift alone with no anti-counterpart
  without affecting the total OAM, which remains constant. These
additional vortices and antivortices thus also display an intricate
dynamics that also affects the morphology, which has to maintain a
constant total and mean angular momentum.  While we do not consider it
here, nonlinearities also result in further vortex-antivortex dynamics
with added complexity to the overall phenomenology.

\section{Harmonic potential}\label{sec:kdfg78t53nejnr}

Next, we consider the displaced vortex in a harmonic potential, so
still with a confinement but that now grows quadratically with the
distance from the center of the potential and thus with the
possibility for the core to stray arbitrarily far from it. We will see
that, surprisingly, such an opportunity is actually seized.
Schr\"{o}dinger's equation $i\hbar\partial_t\psi=H_\mathrm{ho}\psi$ now has
Hamiltonian
\begin{align}
H_\mathrm{ho}=-\frac{\hbar^2}{2m}\nabla^2+\frac12 m \omega_\mathrm{ho}^2 r^2\,,
\end{align}
with $\omega_\mathrm{ho}$ the natural frequency of harmonic
oscillations. Ring potentials in general have been shown to sustain precessing vortices~\cite{Narvarro13,Kev17} with conserved total OAM or even rotating patterns~\cite{Barkhausen20,Kartashov19}, based on combinations of rotational modes of different eigenenergies. The
harmonic potential provides the staple model for many vibrating
systems. Here, we show interesting variations of the vortex fluid
inside such potential in particular regarding the vortex
core. We start our analysis with a displaced-vortex
  that is a spatial equipotential of the trap, that is to say, with a
  width parameter $w=\beta\equiv \sqrt{\hbar/(m\omega_\mathrm{ho})}$ that
  matches the potential's natural frequency:
  \begin{align}\label{eq:jsdhfjd896w978}
    \psi_0=&\frac{e^{-r^2/2w^2}}{w\sqrt{\pi(w^2+r_c^2(0))}}(re^{i\varphi}-r_c(0)e^{i\varphi_c(0)})\,,
  \end{align}
  with
  $\big(x_c(0)\equiv r_c(0)\cos\varphi_c(0),y_c(0)\equiv r_c(0)\sin\varphi_c(0)\big)$
  fixing the initial position of the core. Such a state, that is the
  VB version of a coherent state in the harmonic potential, could be
  implemented in an experiment with two delayed resonant
  pulses~\cite{domicini18a}. The wavepacket dynamics can be obtained
in closed form as
\begin{align}\label{eq:kf83472hfwe84y}
\psi=\frac{e^{-r^2/(2\beta^2)}e^{-i\omega_\mathrm{ho}t}}{\beta\sqrt{\pi(\beta^2+r_c^2(0))}}\big[ re^{i(\varphi-\omega_\mathrm{ho} t)}-r_c(0)e^{i(\varphi_c(0))}\big]\,.
\end{align}

This gives the orbit described by the core, which is a circle of
radius $r_c(0)$. Its motion proceeds with a constant speed
$v_c(0)=r_c(0)\omega_\mathrm{ho}$. Although the core oscillates in time, as
previously and for the same reason, the total angular momentum content
remains constant
\begin{equation}
  \label{eq:hdg7tr23edbjhwgd}
  \langle \tilde{L}_z\rangle=\hbar
  \frac{\beta^2}{\beta^2+r_c^2(0)}=\hbar(1-\frac{r_c^2(0)}{\beta^2+r_c^2(0)})\,,
\end{equation}
and fractional, which is consistent with an offset core. The value
itself depends on the initial radial distance of the core, ranging
from $\langle \tilde{L}_z\rangle=1$ for $r_c(0)=0$ to vanishing values as
$r_c(0)$ goes to infinity. Since the trajectory remains on a circle,
this can be understood as another reason for, or a manifestation of, a
constant angular momentum. In this case where the
  dynamics is smooth and simple, it is interesting to read more deeply
  into Eq.~(\ref{eq:hdg7tr23edbjhwgd}), that one can interpret as the
  total angular momentum arising as a combination of the TC${}=1$ (in
  units of~$\hbar$) and a field contribution
  $-r_c^2(0)/(\beta^2+r_c^2(0))$. Interestingly, since the core itself
  is a clearly identified point (that of no density, or, even more
  clearly in cases of a weak background density, that of a phase
  singularity) with a well-defined position and velocity at all times,
  and since this core itself rotates, one can regard the core as a
  mechanical point with the hope of attaching to it the field
  contribution so that angular momentum is accounted in its entirety
  by the vortex core. To do so, one needs to provide an effective mass
  for the vortex core to define dynamical quantities such as its
  momentum (linear and angular) and its energy (kinetic and
  potential), according to the familiar expressions for a point-like
  particle in terms of the already known position $\mathbf{r}_c$ and
  velocity $\mathbf{v}_c$.  With the following choice for the core's
  effective mass,
  \begin{align}
    \label{eq:346gef87jgjdyefw}
    m_c=-m\frac{\beta^2}{\beta^2+r_c^2(0)}\,,
  \end{align}
  one can check that:
  \begin{subequations}
    \label{eq:Fri14Aug184322CEST2020}
    \begin{align}
      \langle L_z\rangle&=\hbar+m_c(\mathbf{r}_c\times\mathbf{v}_c)\,,\label{eq:Fri14Aug211053CEST2020}\\
      \langle -i\hbar\nabla\rangle&=m_c\mathbf{v}_c\,,\label{eq:Fri14Aug185957CEST2020}\\
      \langle{-\hbar^2\nabla^2\over2m}\rangle&=\hbar\omega_\mathrm{ho}+{1\over2}m_cv_c^2\,,\label{eq:Fri14Aug205009CEST2020}\\
      \langle{1\over2}m\omega_\mathrm{ho}^2r^2\rangle&=\hbar\omega_\mathrm{ho}+{1\over2}m_c\omega_\mathrm{ho}^2r_c^2\label{eq:Fri14Aug205014CEST2020}
    \end{align}
  \end{subequations}
  where the left-hand sides of Eq.~(\ref{eq:Fri14Aug184322CEST2020})
  are obtained as quantum average over the
  wavefunction~(\ref{eq:kf83472hfwe84y}) (with the mass involved $m$
  being that of the field) and the right-hand sides are classical
  expressions for a point-like Newtonian particle (with the mass
  involved being the effective one~$m_c$ for the vortex,
  Eq.~(\ref{eq:346gef87jgjdyefw})). 
  Since Eq.~(\ref{eq:Fri14Aug185957CEST2020}) is two equations (2D
  vector), one single definition of the effective mass fulfills five
  independent dynamical quantities, conferring to this object---the
  vortex core---a clear particle-like character. Note that one
  recovers the quantum results on the lhs by adding quanta to the
  classical expresssions for the core on the rhs. This could be
  expected since the vortex defined as a mechanical point cannot carry
  the vacuum energy of the field, nor the topological charge, which
  are both inherently quantum or wavelike. Interestingly, the
  effective mass is negative and is ``larger'' (in magnitude) the
  closer to the center of the wavepacket, i.e., the more the depletion
  of the density. This explains neatly why the total angular momentum
  is fractional and less than the TC, which appears on its own as one
  quantum of angular momentum~$\hbar$, as expected for a quantum
  vortex. Similarly, both the
  kinetic~(\ref{eq:Fri14Aug205009CEST2020}) and
  potential~(\ref{eq:Fri14Aug205014CEST2020}) energy for the core are
  defined in reference to the vacuum energy~$\hbar\omega_\mathrm{ho}$, which
  the core itself cannot be expected to carry in any classical
  form. As for the linear momentum, which has no intrinsic quantum
  component, the change of sign due to the negative mass can be
  understood as the core going in the opposite direction (in fact
  being radially opposite) to the wavefunction centroid itself, which
  carries the physical angular momentum.

\begin{figure}[tp]
	\begin{center}
		\includegraphics[width=\linewidth]{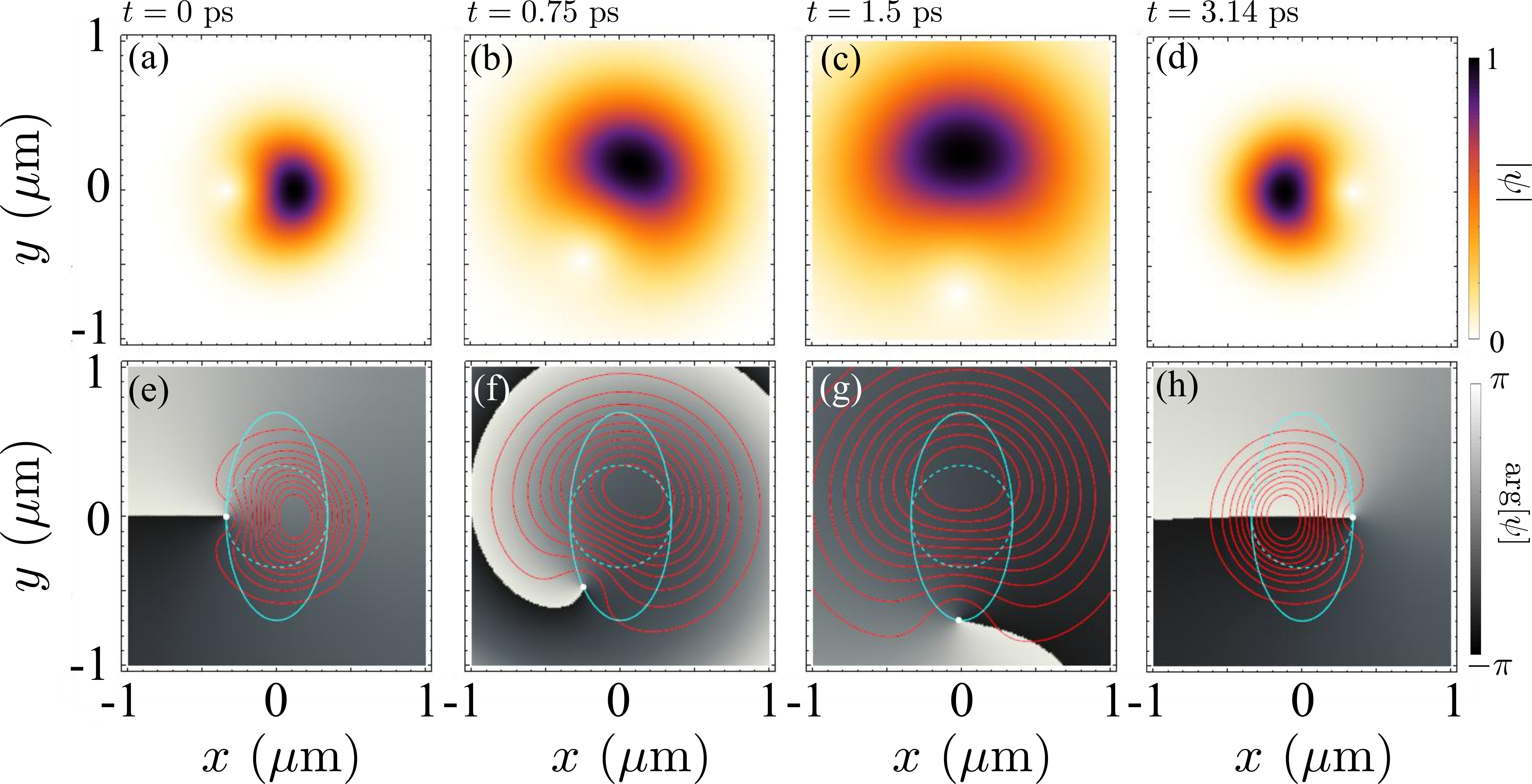}
		\caption{Density plot (upper panels) and phase map (lower panels)
			of a displaced vortex in the non-equipotential case
			$(w=0.7\beta)$ of a symmetric harmonic potential. The trajectory
			of the core is an ellipse shown in solid cyan line (the
			trajectory for the equipotential $(w=\beta)$ case is also shown
			in dashed circle line, for comparison). The vortex morphology is
			shown with red isocontours $|\psi|=$ constant in the phase
			maps.}
		\label{fig:tuwasateitrgywey89382hj}
	\end{center}
\end{figure}

  This connection between quantum and classical pictures through the
  vortex itself also holds when the trajectories are not on
  equipotential lines but involve energy transfers between the kinetic
  and potential terms.  This occurs when the initial condition is not
spatially equipotential in the trap, that is, with $w\neq\beta$. The
solutions are obtained as before
\begin{align}\label{eq:jdfh7379uegiueg}
\psi=\sum_{nm}\alpha_{n,m}(t)\phi_{n,m}(x,y)\,,
\end{align}
where $H_\mathrm{ho}\phi_{n,m}=E_{n,m}\phi_{n,m}$, and
$\alpha_{n,m}(t)=\int e^{-itE_{n.m}/\hbar}\psi_0\phi_{n,m}dxdy$ with
$n,m=0,1,2,\cdots$. The time-evolution of the density is shown in
Fig.~\ref{fig:tuwasateitrgywey89382hj} for $w=0.7\beta$ and for the
core located initially at $(-\beta,0)$. The wavepacket first shrinks
in size, while the core moves but now on an ellipse
$(x_c(t)/w)^2+(wy_c(t)/\beta^2)^2=(\beta/w)^2$ rather than on a
circle, with $\mathbf{r}_c=(x_c,y_c)$ and
$\mathbf{v}_c=d\mathbf{r}_c/dt$ obtained as:
\begin{subequations}
\begin{align}
  \mathbf{r}_c&=-\beta\cos(\omega_\mathrm{ho} t)\hat\imath-\frac{\beta^3}{w^2}\sin(\omega_\mathrm{ho} t)\hat\jmath\,,\\
  \mathbf{v}_c&=\beta\omega_\mathrm{ho}\sin(\omega_\mathrm{ho} t)\hat\imath-\frac{\beta^3\omega_\mathrm{ho}}{w^2}\cos(\omega_\mathrm{ho} t)\hat\jmath\,.
\end{align}
\end{subequations}
Note as well that the speed is now nonuniform. The wavepacket recovers
its initial size as the core completes one period of its motion. Such
oscillations are repeated during the time evolution of the
dynamics. Apart from the peculiar motion of the vortex core, with
accelerations and decelerations along an ellipse, and with also an
intricate time-varying morphology of the phase, the quantum average of
angular momentum is still constant since in general, and for any
initial condition, the expectation value of angular momentum reads as
\begin{align}
\langle L_z\rangle_\mathrm{ho}=i\hbar\sum_{n,m}\alpha_{nm}^{\ast}(0)\big[\alpha_{n-1,m+1}(0)\sqrt{n(m+1)}\nonumber\\-\alpha_{n+1,m-1}(0)\sqrt{m(n+1)}\big]\,,
\end{align}
which is, again, clearly time-independent. This too (like in the case
of the circular quantum well) is expected from the symmetry of the
radial force field (energy gradient) of such a potential, which has a
zero net torque. In its interaction with the fluid, it cannot exert
any change of its overall angular momentum, which therefore
remains. In contrast, the phase morphology and the core, both also
tightly connected to the angular momentum, have a nontrivial time
dynamics. This is satisfied by the core as a mechanical point with now
Keplerian compensation between its distance~$\mathbf{r}_c$ and
velocity~$\mathbf{v}_c$ in their vector product from the right-hand
side of Eq.~(\ref{eq:Fri14Aug211053CEST2020}). Linear
momentum~(\ref{eq:Fri14Aug185957CEST2020}) reads the same and
Eqs.~(\ref{eq:Fri14Aug205009CEST2020}--\ref{eq:Fri14Aug205014CEST2020})
need to have $\hbar\omega_\mathrm{ho}$ upgraded to
$\hbar\omega_\mathrm{ho}\big((\frac{\beta}{w})^\sigma\cos^2(\omega_\text{ho}t)+(\frac{w}{\beta})^\sigma\sin^2(\omega_\text{ho}t)\big)$
with $\sigma=2$ for the kinetic energy and~$\sigma=-2$ for the
potential energy. Such classical/quantum relationships for the core
and its wavefunction could be pushed further but, more interestingly
and for the present discussion, we point out that they break, at least
in this form, when the dynamics of the core becomes clearly
nonclassical.
\begin{figure*}[tb]
	\begin{center}
		\includegraphics[width=.95\linewidth]{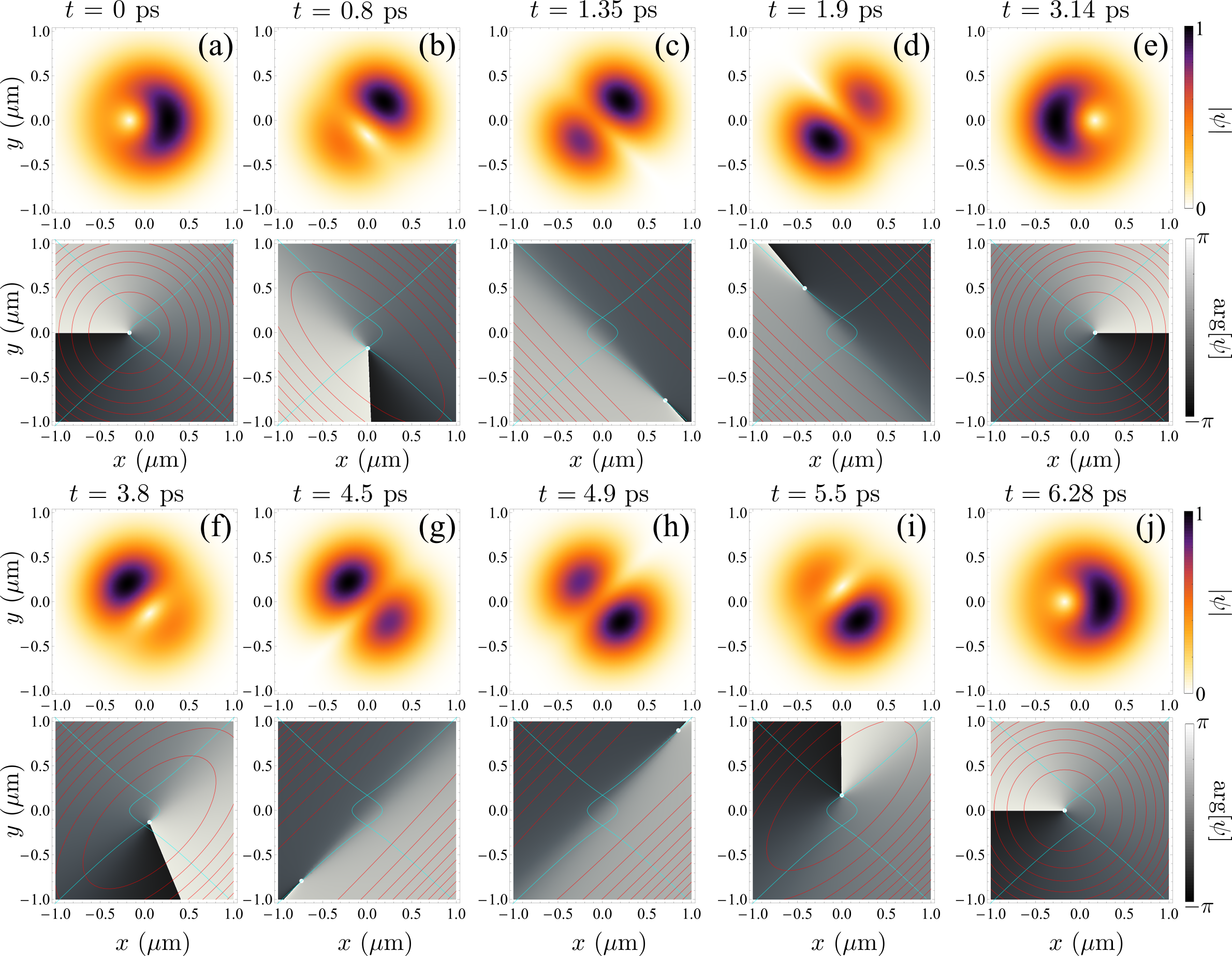}
		\caption{Frames of the dynamics of a displaced-vortex in the
			presence of a quartic anharmonicity in one direction of the 2D
			oscillator, highlighting a case of a VB with time-varying
			angular momentum. The two sets of panels show density and phase
			of the wavefunction at different time frames. The core (white
			dot) moves on a boomerang-shape trajectories (cyan). As it is
			sent to infinity on one branch, it reappears on the other
			branch. Density maps and isodensity contours (red lines) in the
			phase maps show the vortex morphology, which is periodically
			stretched and folded up. The lines evolve from circles at $t=0$
			in (a) to ellipses ((b) and (f)) which get stretched into
			parallel straight lines when angular momentum vanishes and the
			core is sent to infinity. Parameters: $w=\beta$,
			$\omega_\mathrm{ho}=1~\mathrm{ps}^{-1}$.}
		\label{fig:ryteitrgywey8938s2hj}
	\end{center}
\end{figure*}  

To exhibit such a case, we now introduce a perturbation that triggers
a time dynamics of the angular momentum. Namely, we add a spatial
anharmonicity in the form of a quartic contribution added to the
harmonic oscillator potential. Indeed, a fluid in a power law trap
demonstrated rich vortex states including crossover from vortex
lattice to a giant vortex~\cite{Cor11}. We do not consider such
transitions at large rotational velocities here, as we rather focus on
the onset of a time-varying angular momentum due to the anharmonicity,
which could be naturally present in any vibrating system.  Here, for
simplicity, we consider an anisotropic anharmonic term
$H_{an}\equiv \lambda x^4$, with $\lambda>0$. In this case,
coefficients $\alpha_{n,m}$ are given by:
\begin{align}
\alpha_{n,m}(t)=&e^{\frac{-it(E_{n,m}+\lambda \gamma_{n,n})}{\hbar}}\bigg(\alpha_{n,m}(0)\nonumber\\&-i\frac{\lambda}{\hbar}\int_0^t dt' e^{\frac{it'(E_{n,m}+\lambda \gamma_{n,m})}{\hbar}}\sum_{n'\neq n}\alpha_{n,n'}(t')\gamma_{n',n}\bigg)\,,
\end{align}
where $\gamma_{n',n}\equiv\langle
\phi_{n',k}|x^4|\phi_{n,k}\rangle$. Due to the parity of the
eigenstates $\phi_{n,m}$, nonzero values of $\gamma_{n',n}$ are given
for $n+n'$ equal to an even nonzero integer. This simplifies the
computation of $\alpha_{n,m}(t)$. While the angular momentum is
constant for the harmonic potential, even a small deviation, such as
the anisotropic anharmonicity that we have introduced, can result in a
time-varying angular momentum. Clearly, the force field is not central
anymore and can now participate to the rotation of the fluid as a
whole.  Taking as an initial condition $w=\beta$ in
Eq.~(\ref{eq:jsdhfjd896w978}), the anisotropic anharmonicity removes
the degeneracy of the excited states of the harmonic potential, and
even for small $\lambda$, the wavepacket in
Eq.~(\ref{eq:kf83472hfwe84y}) assumes a qualitatively different
form. Namely, it now reads as
\begin{align}\label{eq:sdugfyu7838723}
\psi=&\frac{e^{-r^2/(2\beta^2)}e^{-i(\omega_\mathrm{ho} +\gamma_0)t}}{\beta\sqrt{\pi(\beta^2+r_c^2(0))}}\bigg[e^{-i\omega_\mathrm{ho} t} (x+iye^{-i(\gamma_1-\gamma_0) t})\nonumber\\&-r_c(0)e^{i(\varphi_c(0))}\bigg]\,,
\end{align}
where we have introduced $\gamma_n\equiv \lambda\gamma_{n,n}/\hbar$.
This induces a time-varying angular momentum which reads as:
\begin{align}
\langle \tilde{L}_z\rangle=\hbar \frac{\beta^2}{r_c^2(0)+\beta^2}\cos\big( t(\gamma_1-\gamma_0)\big)\,,
\end{align}
that is, with an angular momentum that not only oscillates in time but
also change signs. The amplitude of oscillations is bounded in a
$(-\hbar,\hbar)$ interval.  The core now evolves on the following
orbits, which correspond to boomerang shapes:
\begin{subequations}
  \begin{align}
    x_c(t)=&r_c(0) \sec\big((\gamma_1-\gamma_0)t\big) \cos\big( (\omega_\mathrm{ho}+\gamma_1-\gamma_0)t+\varphi_c(0)\big)\,,\\
    y_c(t)=&r_c(0)\sec\big((\gamma_1-\gamma_0)t\big) \sin\big( \omega_\mathrm{ho} t+\varphi_c(0)\big)\,,
  \end{align}
\end{subequations}
Examples of the dynamics are shown in
Fig~\ref{fig:ryteitrgywey8938s2hj}. The vortex core (white dot) moves
on two disconnected boomerang trajectories (shown in cyan) as it
shapes the morphology of the vortex phase, which is shown with the red
contours. The most striking departure brought by this case is that the
core now can escape to infinity
(Fig~\ref{fig:ryteitrgywey8938s2hj}a-c), and reappears on another
branch (Fig~\ref{fig:ryteitrgywey8938s2hj}d-g). When the vortex is
infinitely far, the phase morphology is uniform and the total angular
momentum is zero. This thus also corresponds to the point where
angular momentum changes sign. This is a nice illustration of a
complex topological inversion imparted by even a possibly weak
anharmonicity. Such topological inversions have been reported before
with free linear propagation~\cite{molinaterriza01a,zhang19b} and we
hereby provide a counterpart rooted in rotation (and with a potential,
in Section~\ref{sec:jerh98392nf948} we also provide a case in free
space). While the reversal of the total angular momentum in itself is
not a wave-feature of the model, as it is also produced by a classical
oscillator, the behaviour of the core, which is also more striking, is
peculiar to the topological character of the VB. The cycles are also
repeated forever.  Moreover, one can observe an edge dislocation, that
is, when
$t_{\mathrm{cr}}\equiv {\mathcal{N}\pi}/{[2(\gamma_1-\gamma_0)]}$,
with $\mathcal{N}$ an integer. Indeed, the overall movement of the
fluid consists of an additional left and right oscillation due to the
anisotropic term along this direction, in addition to also moving up
and down due to the harmonic trap. This causes dark interference lines
along the diagonal directions, which are reminiscent from the crossing
lines between the two terms of the potential. Notably, these diagonal
edge dislocations are the most pronounced when the vortex core is at
infinity. As pointed out earlier, for a core that exhibits such an
unfamiliar dynamics as far as any mechanical point-like object is
concerned, relationships such as
Eqs.~(\ref{eq:Fri14Aug184322CEST2020}) do not hold anymore, with no
effective mass being able to account simultaneously for all the
Newtonian expressions of the dynamical quantities.
\begin{figure*}[tb]
  \begin{center}
    \includegraphics[width=\linewidth]{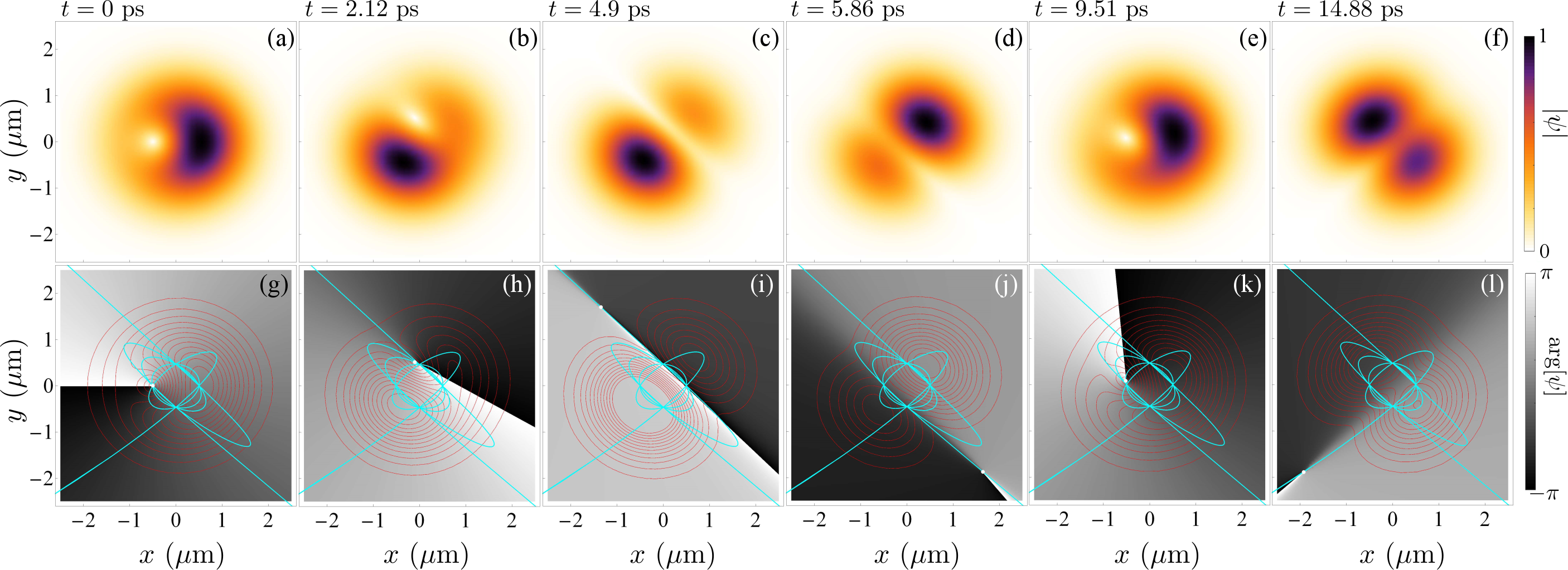}
    \caption{Frames of the dynamics of a displaced vortex in a
      squeezed harmonic potential, with differing confinement in two
      directions. Upper row (a-f) shows the density plot of the field
      ($|\psi|$). Lower row (g-l) shows the phase map of the field,
      with isocontours (in red) of the density. The orbit of the core
      is shown in cyan, with two branches going to
      infinity. Parameters: $\omega_x=2~\mathrm{ps}^{-1}$,
      $\omega_y=2.3~\mathrm{ps}^{-1}$, $x_c=-0.5~\mu\text{m}$, $y_c=0$.}
    \label{fig:ryteitrgywey88s2hj}
  \end{center}
\end{figure*} 

\section{Squeezed Harmonic potential} \label{sec:dbcfweghweywesq}

We now consider a displaced vortex in a squeezed harmonic potential,
where the trap is harmonic, so representing another linear system, but
with a broken symmetry by the trap that arises from its anisotropic
confinement in the two directions. This also gives rise to an
interesting and counter intuitive dynamics of the vortex core, with
time varying angular momentum, similar to the nonlinear case above,
although now in a fully linear system.  The external potential thus
now reads as:
\begin{align}
  H_{\mathrm{SQ}}=-\frac{\hbar^2}{2m}\nabla^2\psi+\frac12 m(\omega_x^2x^2+\omega_y^2y^2)\,,
\end{align}
where $\omega_x\neq\omega_y$ (squeezing) provide the natural
frequencies in the $x$ and $y$ directions, respectively. As
previously, the vortex solutions are found by solving
$i\hbar\partial_t\psi=H_\mathrm{SQ}\psi$, with eigenstates and
eigenenergies of $H_{\mathrm{SQ}}$ being found as
\begin{subequations}
  \label{eq:ghbniueiorjgeoirnknf}
  \begin{align}
    \varphi_{n,m}&=\frac{e^{-((x/\beta_x)^2+(y/\beta_y)^2)/2}}{\sqrt{\beta_x\beta_y}\sqrt{\pi2^{n+m}n!m!}}H_n(x/\beta_x)H_m(y/\beta_y)\,,\\
    E_{nm}&=\hbar(\omega_x(n+\frac12)+\omega_y(m+\frac12))\,,
  \end{align}
\end{subequations}
where $\beta_{x,y}=\sqrt{\hbar/m\omega_{x,y}}$.  The vortex solution
is then expressed as
$\psi=\sum_{n,m}\alpha_{n,m}(0)e^{-i\frac{E_{nm}}{\hbar}t}\varphi_{n,m}(x,y)$
where $\alpha_{n,m}(0)=\int\psi_0(x,y,0)\varphi_{n,m}(x,y)$. Given 
$\psi_0$ as the initial wavepacket for a displaced vortex:
\begin{align}
  \psi_0(x,y,0)=\mathcal{C}e^{-((x/\beta_{x})^2+(y/\beta_{y})^2)/2}\bigg(\frac{x-x_c(0)}{\beta_x}+i\frac{y-y_c(0)}{\beta_y}\bigg)\,,
\end{align}
with
$\mathcal{C}\equiv
\frac{(\beta_{x}\beta_{y})^{-1/2}}{\sqrt{\pi(1+(x_c(0)/\beta_x)^2+(
    y_c(0)/\beta_y)^2)}}$, the subsequent dynamics of the VB is given by
\begin{align}\label{eq:hiuwr7tfguei74udbfiw}
  \psi(x,y,t)=&\mathcal{C}e^{-i(\omega_x+\omega_y) t/2}e^{-((x/\beta_x)^2+( y/\beta_y)^2)/2}\nonumber\\&\times\bigg(\beta_x^{-1}(xe^{-i\omega_x t}-x_c(0))+i\beta_y^{-1} (ye^{-i\omega_y}-y_c(0))\bigg)\,.
\end{align}
It is remarkable that such a trivial-looking solution leads to the
rich and counter-intuitive dynamics that we describe in the
following. We provide an example of it in
Fig~\ref{fig:ryteitrgywey88s2hj}. Initially the vortex core is located
at $(x_c=-0.5,y_c=0)$ and starts to move counterclockwise with a
non-uniform speed. As it accelerates, the core is forced to leave the
density cloud and travels at the periphery of the beam after a few
rotations. Since this is a topological feature, it cannot disappear
altogether even if ejected from the fluid. As was the case for the
anharmonic oscillator, there comes a point in time when the core is
sent away at infinity, in our case along a diagonal in the~$x>0$ part
of the plane, coming back from the opposite quadrant with respect to
the one from which it leaved the plane.  The comeback is done with a
deceleration similar to the acceleration of escape, and as the core
comes back to its initial position, the cycle can repeat, now in the
opposite sense of rotation and escaping along the diagonal of previous
return. Without dissipation, this cyclical dynamics is sustained
forever. The dynamics is also highlighted in a supplementary Movie
SM2~\cite{Movie2}.
The trajectory of the core is obtained from
Eq.~(\ref{eq:hiuwr7tfguei74udbfiw}) as:
\begin{align}\label{eq:jeufr48yreh93wuqpdnbi34}
  \mathbf{r}_c\equiv& \sec(\omega_r t)\bigg[ x_c(0)\cos(\omega_x t)-y_c(0)\sqrt{\frac{\omega_y}{\omega_x}}\sin(\omega_y t)\bigg]\hat{\imath}\,, \nonumber\\
  +&\sec(\omega_r t)\bigg[ y_c(0)\cos(\omega_x t)+x_c(0)\sqrt{\frac{\omega_x}{\omega_y}}\sin(\omega_y t)\bigg]\hat{\jmath}\,,
\end{align}
where an envelope of the form of $\sec(\omega_r t)$ is superimposed to
the otherwise simply elliptical oscillation, and
$\omega_r=\omega_x-\omega_y$. The velocity follows straightforwardly
by taking the time derivative.  As before, this term can bring the
core to infinity, with extreme accelerations.  Depending on the ratio
of $\varsigma\equiv\omega_y/\omega_x$ and the initial position of the
vortex core ($x_c(0),y_c(0)$) the singularity of the secant function
may be removed. For example, when $x_0=0$ and $\varsigma=(2k\pm 1)/2k$
with $k$ as a positive integer, the vortex core does not go to
infinity and stays on a bound orbit, with a stopping point at the end
of each cycle where the motion decelerates and accelerates in the
opposite direction.  The vortex core coordinates themselves,
($x_c(t),y_c(t)$), obey the following equations:
\begin{subequations}\label{eq:kjsieryt934uwiegt}
  \begin{align}
    \frac{d^2x_c}{dt^2}&=\big( -\omega_y^2+\omega_r^2\big)x_c(t)+2\frac{dx_c}{dt}\omega_r\tan\omega_r t\,,\\
    \frac{d^2y_c}{dt^2}&=\big( -\omega_x^2+\omega_r^2\big)y_c(t)+2\frac{dy_c}{dt}\omega_r\tan\omega_r t\,,
  \end{align}
\end{subequations} 
showing how their natural oscillations are at the frequencies
$\omega_{x,y}^2-\omega_r^2$ but with forceful accelerations
($\tan\omega_r t>0$) and decelerations ($\tan\omega_r t<0$) under the
effect of the second term which is proportional to the speed of the
core. This term is induced by the different
frequencies~$\omega_r\neq0$ along the $x$ and $y$ axis, i.e., by the
squeezing of the trap. The time-varying angular momentum can also be
obtained exactly from Eq.~(\ref{eq:hiuwr7tfguei74udbfiw}), as
\begin{align}\label{eq:weuyrf48ryweiuhdf9e}
  \langle \tilde{L}_z\rangle=\frac{\hbar}{2}\pi\mathcal{C}^2(\beta_x^2+\beta_y^2)\cos\omega_r t\,,
\end{align} 
which also makes transparent how the squeezing from the potential
induces time variations of the total angular momentum. Importantly,
the field dynamics is fairly different depending on whether the vortex
is displaced or not. Indeed, for axially symmetric VB ($x_c=y_c=0$),
the density of the field does not rotate, in contrast to the
off-centered case ($x_c\neq0$ and/or $y_c\neq0$) with a clear rotation
of the field, as shown in
Fig.~\ref{fig:ryteitrgywey88s2hj}. Here too, there is
  an interesting topological switching, that is, when
  ${(2\mathcal{N}-1)\pi}<2t\omega_r<{(2\mathcal{N}+1)\pi}$ for
  $\mathcal{N}$ a positive odd integer, the angular momentum of the
  field is positive, while for positive even integers, it is
  negative. Like for the anharmonic case from the previous section,
the core, which even in such a simple system can be brought to extreme
behaviours, cannot be described as a mechanical point-like object with
an effective mass to account by itself for the dynamical properties of
the full field. It remains well-identified as far as its kinematics is
concerned, with an unambiguous position and velocity at all times, but
this does not fit with classical equations of motion. This puts
forward interesting questions on the nature of a vortex and the
conditions sufficient and/or necessary to reduce it to a particle-like
object.

\section{Rabi-coupled fields}\label{sec:jerh98392nf948}

We conclude with the case which inspired all the others considered
previously, since its experimental observation by Dominici \emph{et
  al.}~\cite{domicini18a} triggered our interest into the underlying
mechanism, that we have now generalized to several systems. While we
will focus on polaritons, other similar systems, such as spin-orbit
coupled BEC~\cite{fetter14a}, could also accommodate these
results. Rabi-coupled fields~\cite{RAHMANI2016842} also exhibit
time-varying angular momentum, this time without any external
potential (in free two-dimensional space) but activated for each field
by its coupling to the other. The system is described by a coherent
coupling between two fields $\psi_C$ and $\psi_X$, describing a
cavity-photon field and a quantum-well excitonic field, respectively,
in strong-coupling~\cite{kavokin17a}, with equations that correspond
to two coupled Schr\"odinger equations:
\begin{equation}
\label{eq:Mon22May104821BST2017}
i\hbar\partial_t
\begin{pmatrix}
  \psi_C(x,y,t)\\
  \psi_X(x,y,t)
\end{pmatrix}
=
\mathcal{L}
\begin{pmatrix}
  \psi_C(x,y,t)\\
  \psi_X(x,y,t)
\end{pmatrix}\,,
\end{equation}
where 
\begin{align}\label{eq:Mon22May104821BST2019}
\mathcal{L}=\begin{pmatrix}
-\frac{\hbar^2\nabla^2}{2m_{C}}+E_{C} & \hbar \Omega \\
\hbar \Omega & -\frac{\hbar^2\nabla^2}{2m_{X}}+E_{X}
\end{pmatrix}\,.
\end{align}
Here, $\Omega$ is the Rabi frequency which couples the two fields with
respective free energies~$E_{C,X}$ and masses $m_{C,X}$. There are
typically decay terms to describe such particles which, however, we do
not need to consider for our present discussion, although they also
lead to interesting variations of the dynamics.  As previously, we
assume for the initial condition a displaced core, albeit now with one
vortex in each field:
\begin{align}\label{eq:jsdhfjddsd896w978}
  (\psi_{C,X})|_{t=0}=&\frac{e^{-r^2/2w^2}}{w\sqrt{\pi(w^2+r_{c,x}^2)}}(re^{i\varphi}-r_{c,x}e^{i\varphi_{c,x}})\,.
  % (\psi_X)|_{t=0}=&\frac{e^{-r^2/2w^2}}{w\sqrt{\pi(w^2+r_x^2)}}(re^{i\varphi}-r_xe^{i\varphi_x})\,,
\end{align}
The vortex cores are located at possibly different points in real
space, defined by ($r_{c,x} , \varphi_{c,x}$) in polar coordinates. One can find a closed-form solution in reciprocal space
($k=\sqrt{k_x^2+k_y^2}$ and~$\theta_k=\arg[k_x+ik_y]$) by turning to
the Fourier-transform $\mathcal{F}$ of each field, which yields a
general expression for the amplitudes of the two coupled
fields:
\begin{subequations}
	\label{eq:eyvjahdsgauyd244sada}
	\begin{align}\label{eq:eyvjahdsgauyd244}
	\tilde{\psi}_C=&\frac{1}{k_\Omega^2}e^{-i\frac{(k\lambda)^2M_+}{2\hbar }t}\nonumber\\&\times\bigg[-i\sin(\frac{k_\Omega^2 t}{2\hbar})\left( 2\hbar \Omega (\tilde{\psi}_X)|_{t=0}+(\tilde{\psi}_C)|_{t=0}(\hbar \delta+(k\lambda)^2M_-)\right)\nonumber\\&+(\tilde{\psi}_C)|_{t=0}k_\Omega^2\cos(\frac{k_\Omega^2 t}{2\hbar})\bigg]\,,\\
	\tilde{\psi}_X=&\frac{1}{k_\Omega^2}e^{-i\frac{(k\lambda)^2M_+}{2\hbar }t}\nonumber\\&\times\bigg[-i\sin(\frac{k_\Omega^2 t}{2\hbar})\left( 2\hbar \Omega (\tilde{\psi}_C)|_{t=0}-(\tilde{\psi}_X)|_{t=0}(\hbar\delta+(k\lambda)^2M_-)\right)\nonumber\\&+(\tilde{\psi}_X)|_{t=0}k_\Omega^2\cos(\frac{k_\Omega^2 t}{2\hbar})\bigg]\,,
	\end{align}
\end{subequations}
with $\tilde{\psi}_{C,X}=\mathcal{F}[\psi_{C,X}]$ and where we
introduced
$M_{\pm}\equiv\frac{\hbar \Omega}{2}\left( 1\pm(m_{C}/m_{X})\right)$, $\lambda\equiv\sqrt{\hbar/m_C\Omega}$, and
$k_\Omega^2=\sqrt{(2\hbar \Omega)^2+(\hbar\delta+(k\lambda)^2M_-)^2}$ with
$\hbar \delta\equiv E_C-E_X$ the energy detuning. In general, it is
not possible to find the corresponding closed-form solution in real
space, due to the polaritonic factor~\cite{Rahmani19a}, caused by the
mass imbalance. For equal masses of the exciton and photon fields,
$k_\Omega$ is constant and the dispersion plays no role. One can then
find exact solutions as shown below. These remain excellent
approximations over several periods of the phenomenon when the~$k$
dependence of~$k_\Omega$ is small as compared to the Rabi coupling.
At zero detuning, for instance,
$k_\Omega=(2\hbar)^{(1/4)}\sqrt{\Omega}$ for equal masses and a
similar situation holds for different masses when
$k\ll \sqrt{{4m_C\Omega}/[{\hbar(1-m_C/m_X)}}]$. Since the spread
in~$k$ is related to the packet size~$w$, one can transpose this
condition to read
\begin{equation}
  w\gg\sqrt{\frac{\hbar (1-m_C/m_X)}{2m_C\Omega}}\,.
\end{equation}
In the limit of a very large mass imbalance, as is typically the case,
this yields $w\gg\sqrt{\hbar/(2m_C\Omega)}$, whose right-hand-size
evaluates to about half a micron for standard parameters, showing that
the approximations below are actually enforced by the diffraction
limit for well over the entire duration of the observations for
typical lifetimes of the particles~\cite{domicini18a}.  Therefore,
assuming from now on no effect of the dispersion and taking
$k_\Omega^2\approx \sqrt{(2\hbar\Omega)^2+(\hbar\delta)^2}$, expanding
the expression for the fields~(\ref{eq:eyvjahdsgauyd244sada}) and
considering the initial condition of two displaced vortices
(\ref{eq:jsdhfjddsd896w978}), we find
\begin{subequations}\label{eq:ksdjfsdj0923me}
  \begin{align}\label{eq:jhdg723udcsbybwe7}
    \psi_C\approx&\frac{ we^{\frac{ir^2}{t\Omega'\lambda^2-2iw^2}}}{t\Omega'\lambda^2-2iw^2}\bigg[ \frac{2i\Omega\sin\omega_Rt}{\omega_R \sqrt{\pi(w^2+r_x^2)}}(\frac{2w^2re^{i\varphi
                   }}{t\Omega'\lambda^2-2iw^2}-ir_xe^{i\varphi_x})\nonumber\\&+\frac{ \frac{2w^2re^{i\varphi
    }}{t\Omega'\lambda^2-2iw^2}-ir_ce^{i\varphi_c}}{\sqrt{\pi(w^2+r_c^2)}}\big(\frac{i\delta\sin\omega_Rt}{\omega_R}-2\cos\omega_Rt\big)\bigg]\,,\\
    \label{eq:jwduwgbchwvywdhwb}\psi_X\approx&\frac{ we^{\frac{ir^2}{t\Omega'\lambda^2-2iw^2}}}{t\Omega'\lambda^2-2iw^2}\bigg[ \frac{2i\Omega\sin\omega_Rt}{\omega_R \sqrt{\pi(w^2+r_c^2)}}\frac{ }{}(\frac{2w^2re^{i\varphi
                   }}{t\Omega'\lambda^2-2iw^2}-ir_ce^{i\varphi_c})\nonumber\\&+\frac{ \frac{2w^2re^{i\varphi
    }}{t\Omega'\lambda^2-2iw^2}-ir_xe^{i\varphi_x}}{\sqrt{\pi(w^2+r_x^2)}}\big(\frac{-i\delta\sin\omega_Rt}{\omega_R}-2\cos\omega_R t\big)\bigg]\,,
  \end{align}
\end{subequations}
where $\omega_R\equiv \sqrt{(2\Omega)^2+\delta^2}/2$ is the natural
frequency of the coupled system, and $\Omega'\equiv\Omega(1+m_C/m_X)$.
\begin{figure*}[tp]
  \begin{center}
    \includegraphics[width=\linewidth]{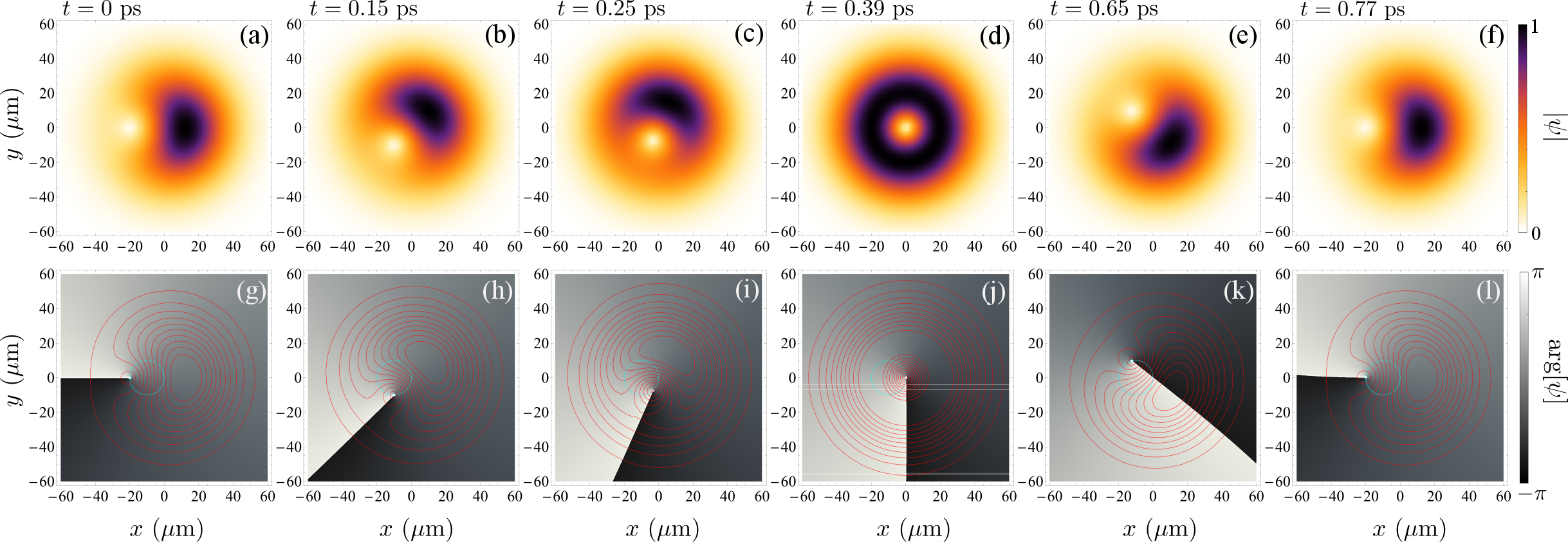}
    \caption{Density plot and phase map in the photon field of a
      Rabi-coupled system with a displaced vortex in each field. The
      core is initially located at $(-w,0)$ and in the limit of
      $m_C/m_X=0.001\ll 1$ for $\psi_C$ in
      Eq.~(\ref{eq:jhdg723udcsbybwe7}). The position of the core is shown
      by a white dot, its trajectory in cyan and the red contours show
      the shape (morphology) of the corresponding vortex beam.}
    \label{fig:tuwasattrgywey89382hj}
  \end{center}
\end{figure*}  
Note that, if initially the two vortex cores are positioned at the
same point in space ($r_c e^{i\varphi_c} = r_x e^{i\varphi_x}$) in
their respective fields, they do not move at all. However, detuning
($\delta\neq0$) can induce phase-driven oscillations of the
density~\cite{Voronova15,Rahmani16,Rahmani19a}, with the morphology
either remaining constant, or otherwise undergoing a complex orbiting
reshaping. The mean total angular momentum per particle for each field is
given as~\footnote{The equation is written by assuming for simplicity in the initial condition that $\varphi_c = \varphi_x$ without losing generality.}:
%
%\begin{align}
%\langle L_z\rangle_{C,X}&=\frac{\hbar w^2}{4}\bigg[ \frac{4\cos^2\omega_R t}{w^2+r_{c,x}^2}+\frac{\sin^2\omega_R t}{\omega_R^2}\big( \frac{(2\Omega)^2}{w^2+r_{x,c}^2}\nonumber\\&\pm\frac{4\Omega\delta}{\sqrt{(w^2+r_c^2)(w^2+r_x^2)}}+\frac{\delta^2}{w^2+r_{c,x}^2}\big)\bigg]\,,
%\end{align}
%
\begin{align}
\langle \tilde{L}_z\rangle_{C,X}&=\frac{\hbar w^2}{4}\bigg[ \frac{4\cos^2\omega_R t}{w^2+r_{c,x}^2}+\frac{\sin^2\omega_R t}{\omega_R^2}\big( \frac{(2\Omega)^2}{w^2+r_{x,c}^2}\nonumber\\&\pm\frac{4\Omega\delta}{\sqrt{(w^2+r_c^2)(w^2+r_x^2)}}+\frac{\delta^2}{w^2+r_{c,x}^2}\big)\bigg]/\nonumber\\&\bigg[\cos^2\omega_R t+\frac{\big(\delta^2\pm\frac{4\delta\Omega(r_cr_x+w^2)}{\sqrt{r_c^2+w^2}\sqrt{r_x^2+w^2}}+4\Omega^2\big)\sin^2\omega_R t}{4\omega_R^2}\bigg]\,,
\end{align}
which is clearly independent of the mass imbalance $m_C/m_X$, and is
also oscillatory in time. Again, it can be observed that for
co-aligned vortices at zero detuning, the momentum stays constant. If
however $\delta\neq0$, even though the cores do not move, there are
some spatial density waves and angular momentum oscillations induced
by the detuning.

In the more general case where the two vortices are not aligned,
thereby exhibiting a relative displacement one to the other, they
exhibit some of the displaced-vortex dynamics of the previous
sections.  The two vortex cores also orbit each other.  Angular
momentum, alongside density, is coherently transferred and oscillating
between the two fields. The net total angular momentum, obtained as
the sum of angular momentum of the two fields, is conserved. The total
angular momentum in each field, however, is time-dependent. Note that
the observation of the system is typically performed over one field
only (the photon field), so this polaritonic mechanism is
intrinsically one that generates periodically time-varying OAM.  It
originates from the exchange of particles with different positions and
momenta~\cite{Voronova15,Rahmani16}, as is well represented by the
cores moving in their off-axis orbits (which are also periodically
changing their distances from the center).  An example of this
polaritonic dynamics is shown in Fig.~\ref{fig:tuwasattrgywey89382hj}
for $m_C/m_X\ll 1$, with frames of the density and phase maps for the
photon ($\psi_C$) field (see also supplementary Movie
  SM3 in~\cite{Movie3}).  The core moves along a circular orbit, shown in cyan, and
the morphology of the vortex, further shown by red isodensity
contours, also displays the type of nontrivial oscillatory structure
similar to those already encountered in the previous systems. Here
too, the core and the fluid as a whole exhibit peculiar and opposed
dynamics, with the core undergoing sequences of accelerations and
decelerations, being faster in the outer part of the
beam~\cite{domicini18a,Rahmani19a}, while the angular momentum
oscillates smoothly and regularly.  The essence of this behaviour can
be reproduced by a simplified version of the coupled field solutions,
namely, in matrix representation (cf.~appendix~\ref{fdguyf873237eug}):
\begin{equation}
\label{eq:Mon2ay104821BST2019}
\begin{pmatrix}
\psi_C\\
\psi_X
\end{pmatrix}
=\mathcal{J}
\begin{pmatrix}
\big(\psi_C\big)|_{t=0}\\
\big(\psi_X\big)|_{t=0}
\end{pmatrix}\,,
\end{equation}
with 
\begin{equation}\label{eq:ufgiuyf89fjsdb}
\mathcal{J}\equiv \begin{pmatrix}
\cos\omega_R t -\frac{i\delta}{2\omega_R}\sin\omega_R t&-\frac{i\Omega}{\omega_R}\sin\omega_R t \\
-\frac{i\Omega}{\omega_R}\sin\omega_R t & \cos\omega_R t +\frac{i\delta}{2\omega_R}\sin\omega_R t
\end{pmatrix}\,.
\end{equation}

Again,
solutions~(\ref{eq:Mon2ay104821BST2019}-\ref{eq:ufgiuyf89fjsdb}) are
valid in the regime when $w\gg1$ only, and would break down for
example, when effects from the dispersion play a
rule~\cite{colas_self-interfering_2016}. In this case further
noteworthy dynamics takes over, with vortex-antivortex pair creation
and recombination similar to the circular quantum well discussed
above, but those are beyond the current scope of this text. The
representation~(\ref{eq:ufgiuyf89fjsdb}) evokes a rotation matrix,
however, it differs due to the complex elements. Such a unitary matrix
is familiar from the dynamics of binary fields in the linear
regime~\cite{Rahmani16} and although it differs from a mere rotation
matrix, it bears an interpretation similar to rotation. To show this,
we introduce a complex number that defines the quantum state in the
basis of the coupled fields $\mathcal{Z}\equiv \psi_C/\psi_X$. The
matrix representation in
Eqs.~(\ref{eq:Mon2ay104821BST2019}-\ref{eq:ufgiuyf89fjsdb}) reappears
as a bilinear transformation
$\mathcal{Z}_0\mapsto \mathcal{Z}=M(\mathcal{Z}_0)$, where
$\mathcal{Z}_0$ is the initial state, with
\begin{align}\label{eq:jhiuedy89yduqwedhuid}
M\equiv \frac{\mathcal{Z}_0\big(\cos\omega_R t -\frac{i\delta}{2\omega_R}\sin\omega_R t\big)-\frac{i\Omega}{\omega_R}\sin\omega_R t}{\mathcal{Z}_0\big(\frac{-i\Omega}{\omega_R}\sin\omega_R t\big)+\cos\omega_R t +\frac{i\delta}{2\omega_R}\sin\omega_R t}\,.
\end{align}
Assuming $\delta=0$ hereafter for simplicity, the transformation
(\ref{eq:jhiuedy89yduqwedhuid}) has two fixed states
$\mathcal{Z}_0=\pm1$, which correspond to the vortex cores in the
dressed states (eigenstates, or normal modes)~\footnote{By definition,
  the dressed states provide a basis of states in which
  (\ref{eq:Mon22May104821BST2019}) is diagonal with hence no evolution
  in time.}. Rewriting the transformation $M$ in the normal form:
\begin{align}
\frac{\mathcal{Z}-1}{\mathcal{Z}+1}=e^{2i\Omega t}\frac{\mathcal{Z}_0-1}{\mathcal{Z}_0+1}
\end{align}
one finds that any $\mathcal{Z}$ is mapped in the complex plane and evolves with a
time-dynamics that keeps it on Apollonius circles~\cite{domicini18a}, which are symmetric
with respect to the fixed points, and that $M$ maps each such circle
in this plane to itself. In advanced complex analysis, the specific
transform $M$ is an example of an elliptic transform~\cite{Tris97}.
This makes $\mathcal{Z}$ points the images in a plane of their
respective location on a Riemann sphere through a stereographic
projection. Denoting a point on the Riemann sphere as $(X,Y,Z)$, the
stereographic expressions is given by:
\begin{subequations}\label{eq:wehdiqwe812juwh}
	\begin{align}
	X+iY=&\frac{2\mathcal{Z}}{1+|\mathcal{Z}|^2}\,,\\
	Z=&\frac{|\mathcal{Z}|^2-1}{|\mathcal{Z}|^2+1}\,.
	\end{align} 
\end{subequations}
The Riemann sphere corresponds to the Bloch sphere of quantum---or
polaritonic---states available to a coupled-system, and it can be
shown~\cite{domicini18a} that the plane can be identified as the
real-physical plane where the fields are evolving. This is revealed by
the motion of the core---neatly identifiable as the single point of
zero density---itself follows such an Apollonius circle.  Bringing the
dynamics on a sphere simplifies it considerably~\cite{Rahmani16}, as
the Rabi oscillations for any quantum state become simply circles on
the sphere (reduced to a point at the poles in the case of polaritonic
eigenstates). The main axis of this sphere is here set horizontal with
respect to the complex plane. Detuning between the modes has the
effect of tilting the plane of the circles on the
sphere. It is worth noting that also the complex
  polar coordinates of the sphere, thanks to the stereographic
  projection, are mapped into two families of mutually orthogonal
  Apollonian circles in the 2D real space as well.  An example of this
  is shown in the Supplementary Movie SM4~\cite{Movie4}. The identification of the
complex plane from the stereographic projection to the physical plane
for the fields allow the interpretation of the two
wavefunctions~$\psi_{C,X}$ as the breakdown of a wider and more
natural object, living in a different space, to describe the
dynamics. This object, the full-wavefunction (as opposed to the
cavity-wavefunction or the exciton-wavefunction), provides the density
of any quantum states present in the system at any time. The peculiar
dynamics observed in Fig.~\ref{fig:tuwasattrgywey89382hj} becomes a
simple rigid rotation in time of the full-wavefunction on its Bloch
sphere, as expected from the linear Rabi dynamics. The role of the
displaced vortices (\ref{eq:jsdhfjddsd896w978}) in this case is to
prepare an initial condition that covers the entire sphere, since the
vortex morphology makes it so that any quantum state is realized at
any time at one, and only one, point in space. Any such point with a
given quantum state undergoes simple Rabi oscillations, corresponding
to, on the sphere, its rotation along its own circle. At such, this
establishes a homeomorphism between the Riemann sphere (Bloch sphere
of quantum states) and the complex plane (real physical
space)~\cite{domicini18a}, that accounts very simply for the vortex
core dynamics, making it only a particular case, namely, the state
perpendicular to that chosen for the observation, in our case, the
pure-excitonic state, since the observation is made with photons. Any
other quantum state can then be seen to undergo a similar
dynamics. This also explains why the core can, in this case like in
previous ones, also be sent arbitrarily far including to infinity, as
a result of the stereographic projection. In all cases, the motion is
a smooth one at uniform speed on the sphere. If the system is prepared
in such a way as to make the corresponding circle on the sphere to
pass by its projection point, the corresponding trajectory will be
distorted in the plane to pass by infinity.  Such a case is considered
in appendix~\ref{fdg873237eug}.  This considerable simplification of
an otherwise intricate dynamics is very appealing and calls for its
generalization to the other systems. The phenomenology being so
similar, one could expect that there also exist equivalent parametric
or phase spaces to be defined in which the intricate dynamics of the
anharmonic or squeezed cases becomes trivial and physically
transparent. This remains for us, however, an open question.

\section{Summary and Conclusions}\label{sec:erk8347yrfn347y}

We have shown how displaced vortices can lead to interesting dynamics
of both the vortex core itself and the total angular momentum of the
field, in a variety of platforms in the linear regime. Our choice of
platforms included various types of confining potentials as well as
coupled condensates (in the limit of low densities where interactions
do not play a role). In all cases, we have highlighted the different
behaviour and character of the vortex as seen through the dynamics of
its core, its morphology (phase map structure) and total angular
momentum. While all are intrinsically connected, they can follow
entirely different types of dynamics, for instance with a complex
underlying evolution of the vortex morphology, with creation of
vortex-antivortex pairs, and cores displaying sequences of
accelerations and decelerations with possible transit to infinity,
while the net angular momentum can remain unaffected. More
specifically, a displaced-vortex in a radial potential keeps its OAM
constant, as expected, despite the core moving on circular or elliptic
orbits with even possible secondary vortex-antivortex pair creations
due to self-interferences of the wavepacket. In the presence of an
asymmetry of the potential, either due to different types or even
simply different magnitudes of the confinement in different
directions, the OAM becomes time-varying and the field displays a
striking phenomenology, notably for the vortex core that can be sent
cyclically to infinity with sequences of extreme accelerations and
decelerations. The phenomenology can also hold without confining
potentials, as illustrated by Rabi-coupled fields where the results
can be further interpreted as a topological homeomorphism linking the
Bloch sphere of possible quantum states for the system with the real
physical plane. The salient result across all the different cases is
that of a striking motion of the vortex core with an intricate
dynamical morphology of the beam, being distinct from and do not
necessarily implying a time-varying OAM per particle, whereas the
latter always imply some offset core and its morphology reshaping.
When this is the case, the core cannot be described as a mechanical
point-like object with an effective mass which otherwise allows it to
account for the dynamical properties of the system, namely, its
momenta (linear and angular) and energies (potential and kinetics).
The absence of interactions was chosen in the models for simplicity
and to provide closed-form solutions, but numerical simulations show
that the phenomenology survives in their presence. This provides
opportunities for new types of micro-control and manipulations of
small objects with exotic wavepackets, similarly to Airy beams
interacting with light particles~\cite{baumgartl08a}. In our case, the
time-varying angular momentum can exert a torque on the objects
immersed in the fluids set in motion as shown in the text, or provide
other services in precision meteorology, such as gyroscopes and even
Casimir torque measurements, or to exploit the periodicity of the
system for instance to realize an hybrid optomechanical torsion
pendulum.  Much control is available from the beam shape and
morphology, that is easily tuneable with a sequence of control optical
pulses, as already demonstrated experimentally. The topology of the
structured light thus emitted could also be useful even in a pure
linear context, offering an additional degree of freedom of possible
benefit for applications of OAM signal encoding and transmission.

%
%\section*{Funding Information}
%
%\section*{Acknowledgments}
\section*{Disclosures}
%
%Disclosures should be listed in a separate section at the end of the manuscript. List the Disclosures codes identified on OSA's \href{http://www.osapublishing.org/submit/review/conflicts-interest-policy.cfm}{Conflict of Interest policy page}. If there are no disclosures, then list ``The authors declare no conflicts of interest.''
%Here are examples of disclosures:
%
%\medskip
%
%\noindent\textbf{Disclosures.} ABC: 123 Corporation (I,E,P), DEF: 456 Corporation (R,S). GHI: 789 Corporation (C).
%
\medskip
The authors declare no conflicts of interest.
\section*{Appendix}
%See supplement 1 for supporting content.
%
\appendix
\section{Simplified coupled-field expressions}\label{fdguyf873237eug}

It is possible to find a simpler version of the solutions given in
Eqs.~(\ref{eq:Mon2ay104821BST2019}--\ref{eq:ufgiuyf89fjsdb}). Starting
from the exact expression for the two coupled fields in the reciprocal
space (\ref{eq:eyvjahdsgauyd244sada}), and ignoring all terms with a
$k$ dependency, the dynamics remains in the real-space domain within
the extent fixed by the size of the initial packet. We can then ignore
diffusive effects or dispersive ones in the dynamics of the
solutions. Note that real-space solutions of (\ref{eq:ksdjfsdj0923me})
include diffusion, which comes from the exponential term
$e^{-i{k^2M_+}t/2\hbar}$ in
Eqs.~(\ref{eq:eyvjahdsgauyd244sada}). Without such effects, one gets
the desired solutions:
\begin{subequations}
  \label{eq:eyvjahdsgauyd24sa4sada}
  \begin{align}\label{eq:eyvjcasdahdsgauyd244}
    \tilde{\psi}_C=&\big[\cos(\omega_Rt)-\frac{i\delta }{2\omega_R}\sin(\omega_Rt)\big](\tilde{\psi}_C)|_{t=0}\nonumber\\&-\frac{i\Omega}{\omega_R}\sin(\omega_Rt)(\tilde{\psi}_X)|_{t=0}\,,\\
    \tilde{\psi}_X=&-\frac{i\Omega}{\omega_R}\sin(\omega_Rt)(\tilde{\psi}_C)|_{t=0}\nonumber\\&+\big[\cos(\omega_Rt)+\frac{i\delta }{2\omega_R}\sin(\omega_Rt)\big](\tilde{\psi}_X)|_{t=0}\,,
  \end{align}
\end{subequations}
which, after taking their inverse Fourier transform, yield the
simplified real-space form (\ref{eq:Mon2ay104821BST2019}).

\section{Vortex core motion on the Riemann sphere}\label{fdg873237eug}

We provide one example of the dynamics of the $\mathcal{Z}$ point on
the Riemann sphere, that involves infinity.  We consider vortex cores
in the photon and exciton fields located at, respectively, ($-w,0$)
and ($0,0$) in real space. We assume no energy-detuning. The
corresponding $\mathcal{Z}$ points are then given as
$\mathcal{Z}_C(t)=i\tan\Omega t$ and
$\mathcal{Z}_X(t)=-i\cot\Omega t$, which move along the imaginary axis
of the complex plane (distinct form 2D real space). At $t=0$, $\mathcal{Z}_X$ is a point at
infinity, in $t=\pi/(2\Omega)$ it reaches the origin of the complex
plane and then in $t=\pi/\Omega$ it is again sent at infinity.  In
contrast, $\mathcal{Z}_C$ is initially positioned at the origin, then
goes to infinity, and finally in $t=\pi/\Omega$ it reappears at the
origin. In both cases, $\frac{d}{dt}\mathcal{Z}_{C,X}$ is not a
constant and such trajectories even involve infinite accelerations and
decelerations.  However, points on the Riemann sphere have a smooth,
uniform-speed dynamics. Indeed, the corresponding points on the sphere
are ($0,-\sin2\Omega t,\cos2\Omega t$) for $\mathcal{Z}_{C}$ and
($0,\sin2\Omega t,-\cos2\Omega t$) for $\mathcal{Z}_{X}$, as is
obtained directly from Eqs.~(\ref{eq:wehdiqwe812juwh}). This is the
equation for a circle on the sphere, on which the points move with a
constant angular speed of $2\Omega$. The described situation can be
realized starting from a specific initial condition of the
form~(\ref{eq:jsdhfjddsd896w978}). The same picture can be extended,
with different parameters but in the same form, to points in the real
space, where the relevant $\mathcal{Z}$ point moves on an associated
Apollonius circle~\cite{domicini18a}.
\section{Captions to Supplementary Movies}\label{fdg87323bsd7eug}

\textbf{Movie SM1} Dynamics of an off-centered vortex in an infinite circular quantum well. Some frames of the movie are shown in Fig.~\ref{fig:rytuwteitrgywey89382hj}.\\
\textbf{Movie SM2} Squeezing the harmonic potential in two directions leads to the weird dynamics of the vortex, that is shown in this Movie. This is for one period of the motion. Examples of the frames are shown in Fig.~\ref{fig:ryteitrgywey88s2hj}. \\
\textbf{Movie SM3} This displays the rotating vortex in Rabi-coupled field in the case of polariton. Here we show the photon field dynamics in one period of the motion. Some frames of photon field ($\psi_C$) are shown in Fig.~\ref{fig:tuwasattrgywey89382hj}.\\
\textbf{Movie SM4} The movie shows the time dynamics of the photon
field (intensity in black, red, yellow, white color scale) undergoing
Rabi oscillations. Here a decay term has been added to illustrate how
the phenomenology evolves in its presence (it produces a tilting of
the sphere and its circles with respect to the plane of projection),
while retaining the main features.  The black circle lines represent
the loci of isophase in the bare modes basis, $\arg(X+iY) = \arg(\mathcal{Z})$=constant, while the white circles
represent the loci of photon-exciton isocontent, $Z$ = constant. The nodal points to the isophase circles
represent the position of the vortices in the two bare modes.
\bibliography{ta,Sci}
%\bibliographyfullrefs{ta}
%
\end{document}